\documentclass[preprint2]{aastex}

\newcommand{\jh}{\textit{J--H}}
\newcommand{\hk}{\textit{H--K}}
\newcommand{\ji}{\textit{J}}
\newcommand{\hi}{\textit{H}}
\newcommand{\ki}{\textit{K}}

\newcommand{\ksi}{\textit{K}\ensuremath{_{s}}}
\newcommand{\av}{\ensuremath{A_{\rm V}}}

\newcommand{\fx}{\ensuremath{F_{X}}}

\slugcomment{submitted to AJ}
\shorttitle{Deep NIR Observations on the OMC-2/3}
\shortauthors{Tsujimoto et al.}

\begin{document}
\title{Deep Near-Infrared Observations and Identifications of \textit{Chandra} Sources in the Orion Molecular Cloud 2 and 3}
\author{Masahiro Tsujimoto, and Katsuji Koyama}
\affil{Department of Physics, Graduate School of Science, Kyoto University, Sakyo-ku, Kyoto, 606-8502, Japan}
\email{tsujimot@cr.scphys.kyoto-u.ac.jp, koyama@cr.scphys.kyoto-u.ac.jp}
\author{Naoto Kobayashi, and Miwa Goto\altaffilmark{1}}
\affil{National Astronomical Observatory of Japan, 650 North A'ohoku Place, Hilo, HI 96720, USA}
\email{naoto@subaru.naoj.org, mgoto@subaru.naoj.org}
\altaffiltext{1}{Visiting Astronomer at the Institute for Astronomy, University of Hawaii}
\author{Yohko Tsuboi}
\affil{Department of Physics, Faculty of Science and Engineering, Chuo University, Bunkyo-ku, Tokyo, 112-8551, Japan}
\email{tsuboi@phys.chuo-u.ac.jp}
\and
\author{A. T. Tokunaga}
\affil{Institute for Astronomy, University of Hawaii, 2680 Woodlawn Drive, Honolulu, HI 96822, USA} 
\email{tokunaga@ifa.hawaii.edu}

\begin{abstract}
We conducted deep near-infrared (NIR) imaging observations of the Orion molecular cloud
 2 and 3 using Quick Infrared Camera on the 88-inch telescope of the University of
 Hawaii. Our purposes are 1)~to generate a comprehensive NIR source catalog of these
 star forming clouds, and 2)~to identify the NIR counterpart of the 
 \textit{Chandra} X-ray sources that have no counterpart in the Two Micron All Sky
 Survey (2MASS) catalog. Our \ji-, \hi-, and \ki-band observations are $\sim$2~mag deeper
 than those of 2MASS, and well match the current \textit{Chandra} observation. We
 detected 1448 NIR sources, for which we derived the position, the \ji-, \hi-, and
 \ki-band magnitude, and the 2MASS counterpart. Using this catalog, we identified the
 NIR counterpart for $\sim$42\% of the 2MASS-unIDed \textit{Chandra} sources. The nature
 of these \textit{Chandra} sources are discussed using their NIR colors and spatial
 distributions, and a dozen protostar and brown dwarf candidates are identified.
\end{abstract}

\section{Introduction}
The \textit{Chandra X-ray Observatory} made X-ray observations to be
one of the most powerful techniques for the study of star formation. Its sensitivity in
the hard X-ray band ($E>$2~keV) provides us a new light to probe into the dense cloud
cores of \av$\gtrsim$100~mag extinction, and its spatial resolution of $\sim$ 1\arcsec\
can separate virtually all the individual cloud members. In addition, its very low
detector noise enables us to access very faint X-ray sources like deeply embedded
protostars and brown dwarfs, which were quite difficult with the preceding X-ray
observatories.

\citet{casanova95} derived an empirical relation between the un-reddened \ji-band
magnitude (\ji) and the X-ray flux (\fx) in the 1.0--2.4~keV band using T Tauri star
samples in the $\rho$ Ophiuchi cloud ($D=$160~pc) detected by \textit{ROSAT}, which is
\begin{equation}
 \log{F_X (4\pi D)^2} \sim -0.30 J + 32\label{eq:e1}.
\end{equation}
The detection limit of the \textit{Chandra} observation of the Orion nebula cluster, for
instance, is
\fx$\sim$4$\times$10$^{-16}$~erg~s$^{-1}$~cm$^{-2}$ in the 0.5--8.0~keV band
\citep{feigelson02}, which can be converted to
\fx$\sim$2$\times$10$^{-16}$~erg~s$^{-1}$~cm$^{-2}$ in the 1.0--2.4~keV band assuming
a thin-thermal plasma spectrum of 1~keV temperature. If equation (\ref{eq:e1}) can
be applied to other young stellar objects (YSOs), the required near-infrared (NIR)
detection limit to match the \textit{Chandra} depth would be \ji$\sim$17~mag. The
deepest all-sky NIR catalog is the Two Micron All Sky Survey (2MASS) down to
\ji$\sim$15.8~mag\footnote{See http://www.ipac.caltech.edu/2mass/.}. In some star
forming regions, therefore, X-ray observations are ``deeper'' than NIR observations.

The Orion molecular cloud 2 and 3~(OMC-2/3), a member of the Orion--Monoceros
complex \citep{maddalena86}, is one of these regions. \citet{tsujimoto02a} detected 385
X-ray sources with the ACIS-I onboard \textit{Chandra}, and 203 were identified to have
the 2MASS \ji-, \hi-, and \ksi-band counterpart. They discussed the natures of these X-ray
sources based on their NIR colors. However, the natures of X-ray sources with no 2MASS
counterpart (153 sources) or no 2MASS \ji- and/or \hi-band detections (29 sources)
remain unknown. Most of these 182 X-ray sources (hereafter we call them ``2MASS-unIDed''
sources) are  faint, hence some may be protostars and/or brown dwarfs. In order to
investigate the nature of 2MASS-unIDed sources, NIR imaging observations of 2--3~mag
deeper than the all-sky survey depth are necessary.

Here, we report the result of our deep \ji-, \hi-, \and \ki-band observations of
OMC-2/3. This paper has two purposes; 1)~to generate a comprehensive NIR source catalog
of this region, and 2)~to identify the NIR counterpart of the 2MASS-unIDed
\textit{Chandra} sources and to discuss their nature based on their NIR colors and
spatial distributions, with the emphasis on the search for protostar and brown dwarf
candidates. The X-ray features of these candidates will be discussed in a separate paper. 

The OMC-2/3 is known to have the highest condensation of protostars among the nearby
star forming clouds \citep{reipurth99}, and are superb fields to study the initial
conditions of star formation. Extensive observations have been reported particularly 
in the millimeter \citep{bally87,tatematsu93,dutrey93,cesaroni94,castets95,chini97,yu00,aso00} to
sub-millimeter \citep{lis98,johnstone99} band, detecting 20--30 protostellar cores. In
the NIR band, however, previous studies
\citep{gatley74,thronson82,pendleton86,johnson90,jones94,ali95,carpenter00,carpenter01b}
were all confined to OMC-2 and/or not deep enough (particularly in the \ji\ band) to
pick up deeply embedded protostars. Since sensitive \ji-, \hi-, and \ki-band imaging
observations are the most efficient and common technique to detect YSOs in protostellar
cores, our imaging survey would also be helpful in search for embedded YSOs in these
protostellar cores.

Throughout this paper, we assume the distance to OMC-2/3 to be 450~pc \citep{genzel89}.

\section{Observation}
We used the Quick Infrared Camera (QUIRC) mounted on the Cassegrain focus of the 88 inch
telescope of the University of Hawaii \citep{hodapp97}. QUIRC is a 1024$\times$1024-pixel
HgCdTe array, providing a 3\farcm2$\times$3\farcm2 field of view (FOV) with the pixel
scale of 0\farcs189. The smaller pixel scale than 2MASS detectors are more suitable
to pick up NIR sources contaminated by diffuse emissions, particularly at the southern
half of our studied field.

\begin{figure}
 \figurenum{1}
 \epsscale{1.0}
 \plotone{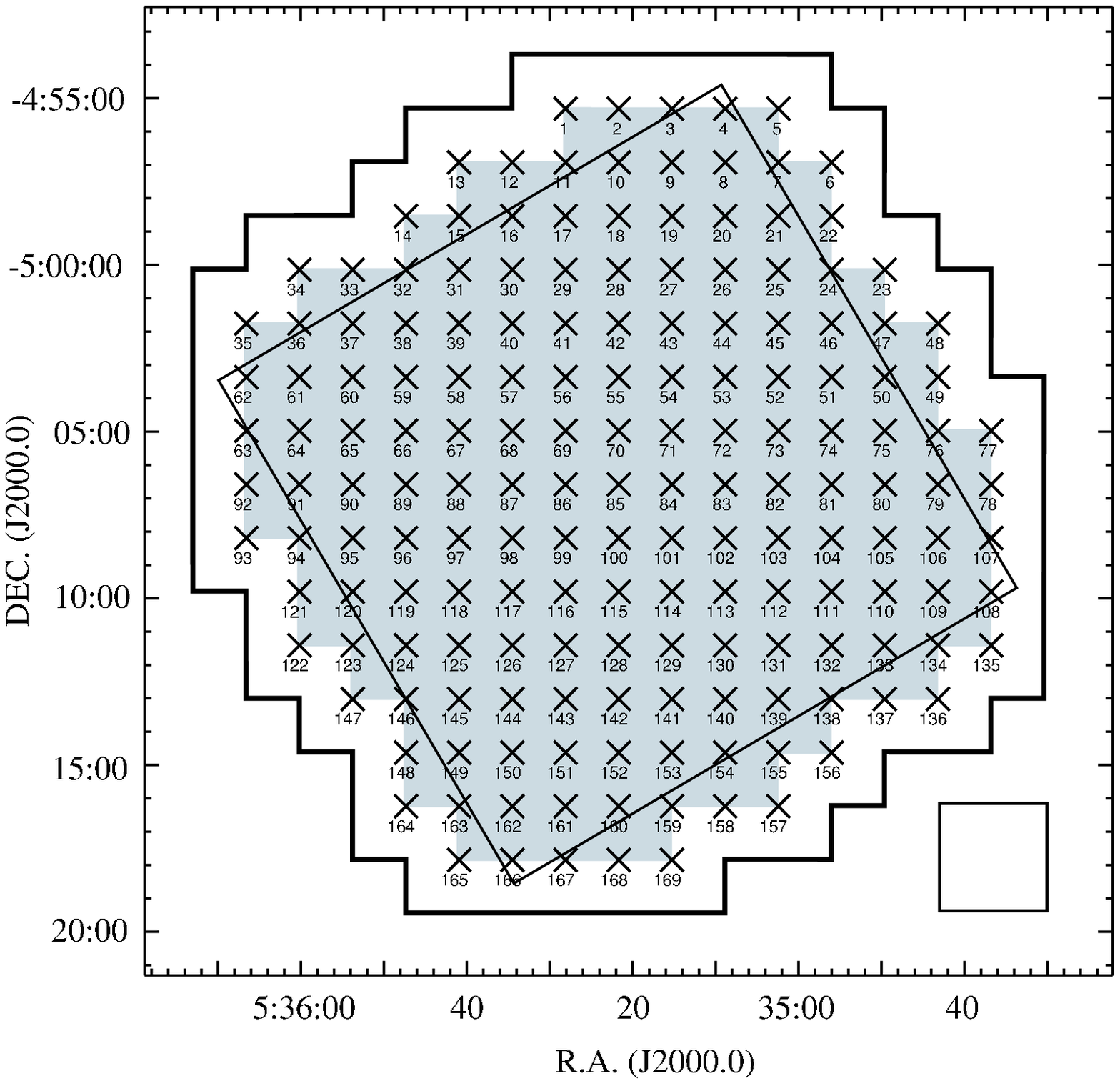}
 \caption{The configuration of our mosaic mapping. The positions of 169 frame centers
 are marked in crosses with the relevant frame number. The frames are tiled at
 the intervals of 1\farcm6 to sweep the \textit{Chandra} field (the solid oblique
 square). Each frame covers the 3\farcs2$\times$3\farcs2\ square region (the size is
 shown at the right bottom) centered at each cross. When all frames combined, any point
 in the thick lines ($\sim$512~arcmin$^{2}$) is covered at least by one frame, while
 that in the gray region ($\sim$360~arcmin$^{2}$) is covered by four
 frames.}\label{fg:f1}
\end{figure}

We conducted mosaic mapping observations to sweep the whole \textit{Chandra} field
(Fig.~\ref{fg:f1}). We first covered the uppermost row (filed No.1--No.5) of the
mosaic by shifting the field center by 1\farcm6 (half of the FOV of QUIRC) from
east to west. Then, we moved 1\farcm6 southward and covered the second uppermost row
(field No.6--No.13) by inversely shifting from west to east. We continued these
raster scans to reach the lowermost row of the mosaic. 
In this way, we can minimize the dead time due to the shift of the field center, and
reduce the effect of ghosts. In total, we swept the \textit{Chandra} FOV with 169 QUIRC
FOVs that amount to $\sim$512~arcmin$^{2}$.

Each 169 frames was exposed for 60~s. We conducted the mosaic mapping once in the
\hi\ and \ki\ band, and twice for the \ji\ band, spending six half nights of 2001
February 4--6 and 2001 March 11--13 (Table \ref{tb:t1}). All nights were photometric, and
the seeing was 0\farcs7--1\farcs1. In one sweep, any region inside the \textit{Chandra}
field except for edges (the gray region in Fig.~\ref{fg:f1}) was covered four times, so
the nominal exposure time was 240~s in the \hi\ and \ki\ band, and 480~s in the \ji\ band.

The studied field covers the whole of OMC-2 and OMC-3 clouds (Fig.~\ref{fg:f2}).

\section{Analysis}
All the QUIRC frames were reduced following the standard procedures using IRAF\footnote{IRAF
is distributed by the National Optical Astronomy Observatories, which are operated by
the Association of Universities for Research in Astronomy, Inc., under cooperative
agreement with the National Science Foundation.}; i.e., dark-subtraction,
flat-fielding, sky-subtraction, and bad pixel removal. Flat frames were taken with the
dome flat, while sky frames were constructed by the median sky technique for each band
and each night. SExtractor \citep{bertin96} was used for source extraction and
photometry.

For the purpose of the astrometric and photometric calibration of all frames, we
consulted the Second Incremental Data Release of 2MASS catalog. This catalog provides
the position and the \ji-, \hi-, and \ksi-band magnitudes of NIR sources down to 15.8,
15.1, and 14.3~mag. The accuracy in position and magnitude is $\sim$0\farcs1 and
$\sim$0.1~mag in the range of 6--15~mag.

We tentatively extracted sources from each QUIRC frame and correlated them with 2MASS
sources using WCSTools\footnote{See 
http://tdc-www.harvard.edu/software/wcstools/.}. All
frames were found to have 18--145 sources, including 4--51 sources with the 2MASS
counterpart. Using the QUIRC--2MASS counterpart pairs, we first shifted each QUIRC frame
so that the mean separation between QUIRC sources and their 2MASS counterpart reaches
the minimum. Second, we multiplied each QUIRC frame with a constant value
that was derived by the least square method in order to match the QUIRC photometry with
the 2MASS photometry. In this way, we used the 2MASS sources as standard stars. Sources
brighter than 11~mag were not used for this procedure because of the non-linearity of
QUIRC (see Sect.~\ref{sec:4-3}), and those fainter than 16~mag were not used either
because of the large uncertainty of the 2MASS photometry data.

All the frames, which were thus corrected for astrometry and photometry, were combined
into three large mosaic images of the \ji, \hi, and \ki\ band.

\section{Results}
\subsection{Source Extraction and Photometry}
In prior to the source detection, all the images were binned with the neighboring
2$\times$2 pixels and smoothed with a gaussian function to attain better signal-to-noise
ratio. For the \ki-band mosaic image thus improved, we extracted NIR sources above
3$\sigma$ level. In SExtractor, we can choose any convolution masks in the peak finding
procedure. We adopted a mexican hat function with the seeing size radius for the
mask. This function gave the most robust result among several masks with various radii
we examined, particularly in regions contaminated with diffuse emissions. With visual
inspections of the output, we carefully removed 1)~sources at the edge of the mosaic
images, 2)~ghosts of bright sources, and 3)~spurious detections (in most cases, diffuse
structures were identified as point-like sources). As a result, we picked up 1448
\ki-band sources.

For each \ki-band detected source, we derived the \ji-, \hi-, and \ki-band magnitude
with the adaptive aperture photometry implemented in SExtractor. For source less than
3$\sigma$ detection in the \ji\ and/or \hi\ band, we calculated the 3$\sigma$ upper
limit of their magnitudes.

We correlated all the QUIRC sources with the 2MASS catalog, and found that 682
($\sim$47\%) have a 2MASS counterpart.

\subsection{Source List}
Table \ref{tb:t2} lists the QUIRC sources with their source number, position, the \ji-,
\hi-, and \ki-band magnitudes, and their 2MASS counterpart. Hereafter, we follow the IAU
convention for the source naming; i.e., TKK~J05342894$-$0508387 for the source No.1. The
label ``$>$'' means the 3$\sigma$ upper limit of \ji- and/or \hi-band fluxes (lower
limit of the magnitude). For the saturated QUIRC sources (brighter than 11~mag; see
Sect.~\ref{sec:4-3}), we replaced them with the 2MASS magnitudes with the label
``$\dagger$'' if available in the 2MASS catalog. If not, we labeled the magnitudes with
``$<$'' to show the lower limit of the flux (the upper limit of the magnitude).

\subsection{Astrometric and Photometric Accuracy}\label{sec:4-3}
\begin{figure}
 \figurenum{3}
 \epsscale{1.0}
 \plotone{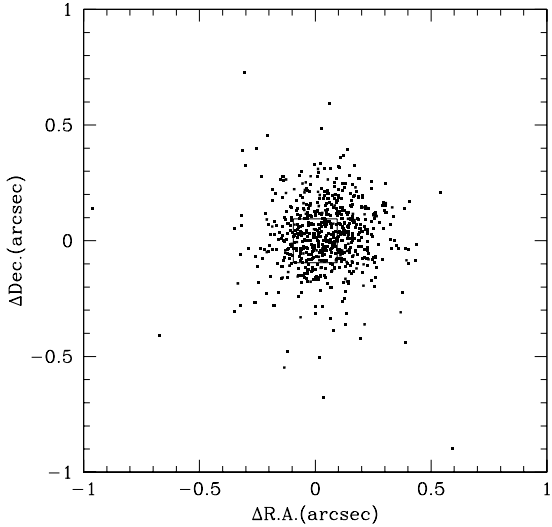}
 \caption{The astrometric accuracy of QUIRC sources. The difference of R.A. and
 Dec. ($\Delta$R.A. and $\Delta$Dec.) between the QUIRC and 2MASS positions are plotted
 for QUIRC--2MASS pairs. The solid square at the center is the QUIRC pixel scale of
 0\farcs189$\times$0\farcs189.}\label{fg:f3}
\end{figure}

Using the QUIRC--2MASS counterpart pairs, we evaluated the accuracy of the QUIRC
astrometry and photometry in the following way. In Fig.~\ref{fg:f3}, we plotted
differences between the QUIRC and 2MASS positions both in the right ascension (R.A.) and
the declination (Dec.). We found that 
$\Delta$R.A.$=0\farcs048\pm0\farcs161$ (1$\sigma$)
and $\Delta$Dec.$=0\farcs022\pm0\farcs148$ (1$\sigma$), which indicates that the QUIRC
positions are determined with the accuracy of the QUIRC pixel scale (0\farcs189) and the
systematic offset between the QUIRC and 2MASS positions is negligible.

\begin{figure}
 \figurenum{4}
 \epsscale{1.0}
 \plotone{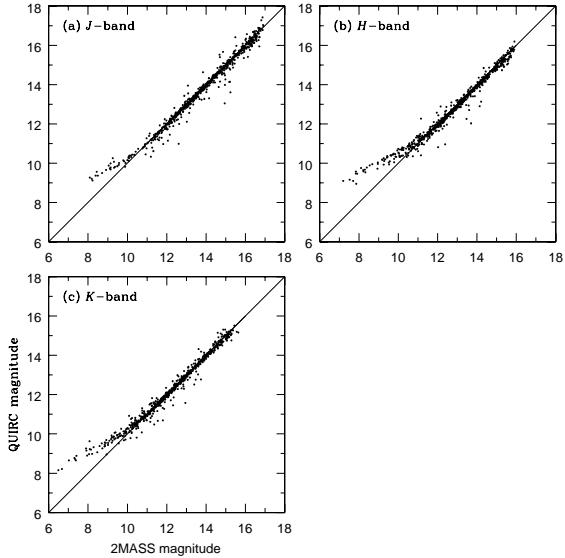}
 \caption{The photometric accuracy of QUIRC sources. The 2MASS and QUIRC magnitude of
 QUIRC--2MASS counterpart pairs are plotted separately for (a) \ji\ band, (b) \hi\ band,
 and (c) \ki\ band. Solid lines represent (2MASS magnitude$=$QUIRC
 magnitude).}\label{fg:f4}
\end{figure}

In Fig.~\ref{fg:f4}, we plotted the QUIRC and 2MASS magnitudes of QUIRC--2MASS
counterpart pairs separately for each band. The linear relation (2MASS magnitude $=$
QUIRC magnitude) is violated at the brighter side than $\sim$11~mag due to the
non-linearity of the QUIRC detector. The QUIRC saturation limit of 44,000 ADU counts
(Hodapp~et~al. 1997) corresponds to $\sim$11~mag in our observation. For sources fainter
than 11~mag in the QUIRC and 2MASS magnitude, we calculated the difference between their
magnitude for each band ($\Delta m_{J}$, $\Delta m_{H}$, and $\Delta m_{K}$). We found
that $\Delta m_{J}=-0.06\pm0.22$~mag (1$\sigma$), $\Delta m_{H}=-0.07\pm0.19$~mag
(1$\sigma$), and $\Delta m_{K}=-0.04\pm0.18$~mag (1$\sigma$), indicating that the QUIRC
photometry fainter than 11~mag is reliable within $\sim$0.2~mag of the 2MASS photometry.

\subsection{Survey Depth}
\begin{figure}
 \figurenum{5}
 \epsscale{1.0}
 \plotone{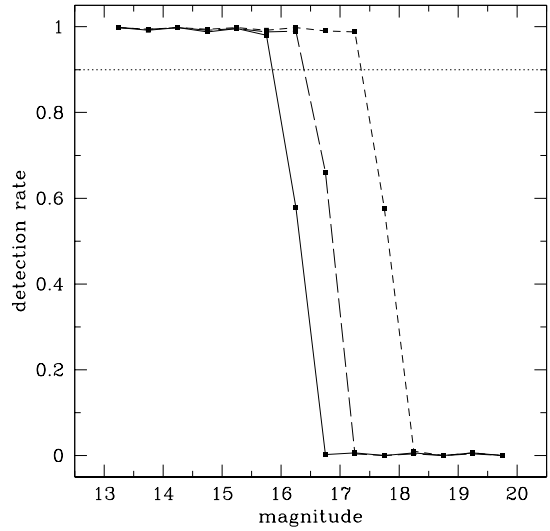}
 \caption{The completeness limit of QUIRC observations. The fraction of detected
 artificial sources is shown with solid (\ki\ band), long-dashed (\hi\ band), and the
 short-dashed (\ji\ band) lines. The 90\% (the dotted line) completeness limit is
 \ki$\sim$16.0~mag, \hi$\sim$16.5~mag, and \ji$\sim$17.5~mag.}\label{fg:f5}
\end{figure}

We estimated the survey depth of our QUIRC observations in the following manner. First,
we embedded 500 artificial sources with 13.0--13.5~mag in the \ji-, \hi-, and \ki-band
mosaic image. The same source detection algorithm was employed to detect these
artificial sources, then the detection rate of sources with 13.0--13.5~mag was
derived. The same procedure was repeated for sources of different magnitudes from
13.0--20.0~mag with 0.5~mag bins. The detection rate at each magnitude bin is given in
Fig.~\ref{fg:f5} for the \ki\ (solid), \hi\ (long-dashed), and \ji\ (short-dashed) band
respectively. The 90\% completeness limit was thus estimated to be \ki$\sim$16.0~mag,
\hi$\sim$16.5~mag, and  \ji$\sim$17.5~mag. 

\begin{figure}
 \figurenum{6}
 \epsscale{1.0}
 \plotone{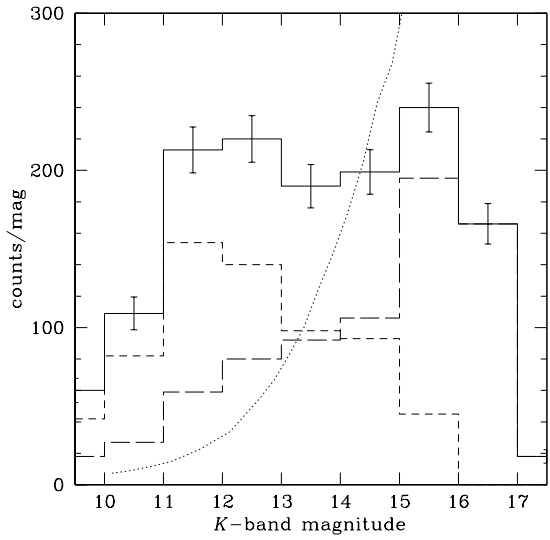}
 \caption{The QUIRC \ki-band-detected source counts ($N(M)$) at each magnitude
 ($M$). The solid histogram is the counts of all the QUIRC \ki-band sources, while the
 short- and long-dashed histograms are those with and without the 2MASS counterpart,
 respectively. The uncertainty of $\sqrt{N(M)}$ is given for each bin of the solid
 histogram. The dotted curve shows the back- and fore-ground source counts predicted by
 the Galactic star count model assuming no extinction. This curve should be shifted
 rightward corresponding to an average extinction of 1--2~mag.}\label{fg:f6}
\end{figure}

In order to compare the depth of our observations with that of 2MASS, we made the
histogram of the number counts of the \ki-band detected source at each magnitude (\ki-band
luminosity function; KLF) in Fig.~\ref{fg:f6}. The short- and long-dashed histograms
respectively represent the KLF of QUIRC sources with and without the 2MASS counterpart,
while the total is given in the solid histogram. The 2MASS catalog fails to detect some
bright sources even in \ki$\lesssim$14~mag, which is mainly due to the contamination by
diffuse emission or the binary companion. This can be confirmed by plotting the 2MASS
source list on the 2MASS image.

\subsection{Cloud Membership and Age}
In the KLF shown in Fig.~\ref{fg:f6}, we can estimate the back- and fore-ground source
contamination.  The dotted curve shows the back- and fore-ground source counts predicted
by SKY; a Galactic star count model \citep{cohen93,cohen94,cohen94b,cohen95} assuming no
extinction. We examined colors of all the detected sources and found that they have an
average extinction of $A_{K}\sim$1--2~mag, which shifts the dotted line rightward
correspondingly. The contamination is not negligible for sources fainter than
\ki$\sim$15~mag, but for brighter sources, we can assume that most of them are cloud
members.

The peak of KLF at 12--13~mag is a real feature when the completeness limit and back-
and fore-ground contamination are taken into consideration. This has been inferred by
\citet{jones94}, who studied the KLF of OMC-2 sources. The peak of KLF is often seen in
the young associations like Orion Nebula Cluster \citep{ali95,hillenbrand00}, and is
considered to be caused by the difference of the mass-to-luminosity relation between
main-sequence and pre-main-sequence stars \citep{muench00}. The peak is consistent with
the cloud age of $\sim$1~Myr \citep{johnson90} that was derived by plotting some bright
NIR sources in OMC-2 on the HR diagram, and 10$^{6\pm0.4}$~yr \citep{rebull00} that was
estimated from the \textit{I}- and \textit{V}-band magnitudes of Orion sources.

\section{\textit{Chandra} Sources with no 2MASS Counterpart (2MASS-unIDed Sources)}
\subsection{QUIRC Counterpart for 2MASS-unIDed Sources}
Using the QUIRC source list (Table \ref{tb:t2}), we searched for the NIR counterpart of
182 2MASS-unIDed sources. The radius of the point spread function (PSF) of
\textit{Chandra} differs by more than an order of magnitude between on-axis sources and
those near the field edge, which deteriorates the position accuracy at large off-axis
sources. Therefore, in identifying the QUIRC counterpart of the \textit{Chandra}
sources, we took the off-axis angle into account in the following manner.

\begin{figure}
 \figurenum{7}
 \epsscale{1.0}
 \plotone{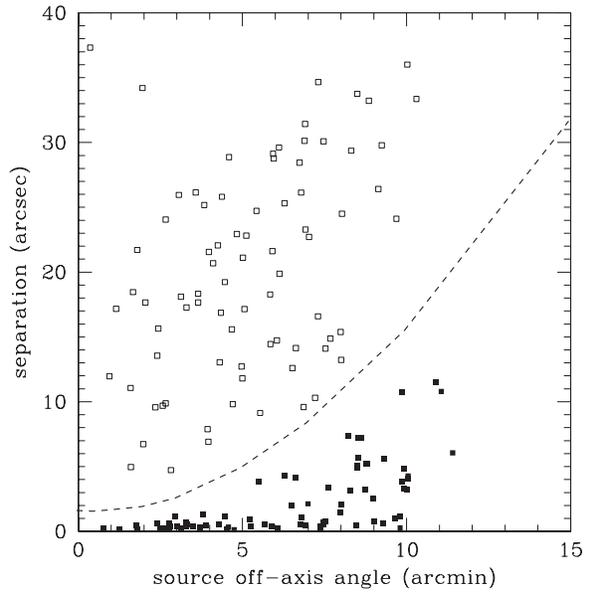}
 \caption{The closest QUIRC--\textit{Chandra} pairs are plotted with their separation
 and the \textit{Chandra} off-axis angle on the vertical and the horizontal axis,
 respectively. The dashed curve is the 90\% encircled energy radius of 1.49~keV X-rays
 at a given off-axis angle. Filled squares, which are below the dashed line are
 recognized as the counterpart pairs, while open squares above the dashed line are
 recognized as non-associated pairs.}\label{fg:f7}
\end{figure}

First, we searched for the QUIRC source closest to each \textit{Chandra} source. Second,
we conversely searched for the \textit{Chandra} source closest to each QUIRC
source. Then, we picked up 158 QUIRC--\textit{Chandra} pairs that are the closest to
each other. These include both pairs of physically associated (``counterpart
pairs'') and of no physical association (``non-associated pairs''). In
Fig.~\ref{fg:f7}, we plotted the separation between the closest QUIRC--\textit{Chandra}
pairs as a function of the \textit{Chandra} off-axis angle. We see two groups (filled
and open squares), which are well separated by the dashed line
indicating the 90\% encircled energy radius ($r_{90}$; $\sim$ 0.9$\times$FWHM of the
gaussian PSF) of 1.49~keV X-rays as a function of off-axis angle\footnote{See
http://asc.harvard.edu/udocs/docs/POG/MPOG/index.html.}. Since $r_{90}$ is
the radius in which the 90\% of incident X-ray photons are accumulated, it also
represents the position accuracy of the \textit{Chandra} sources at each off-axis
angle. We therefore regarded the closest QUIRC--\textit{Chandra} pairs with the separation
angle less than $r_{90}$ (filled squares) to be the counterpart pairs, and those with larger
separation angle (open squares) as the non-associated pairs. As a result, 74
2MASS-unIDed sources were newly found to have a QUIRC counterpart
(hereafter we call them ``QUIRC-IDed'' sources), which are listed in Table \ref{tb:t3}.

\subsection{The Nature of QUIRC-IDed {\textit{Chandra}  Sources}}\label{sec:5-2}
For the QUIRC-IDed sources, we estimated their nature using their NIR
colors, putting particular interest on X-ray emitting protostars and brown dwarfs. We
converted the \ji-, \hi-, and \ki-band magnitudes in Table \ref{tb:t2} (\ji$_{\rm{QUIRC}}$,
\hi$_{\rm{QUIRC}}$, and \ki$_{\rm{QUIRC}}$) to the California Institute of Technology
(CIT) color system (\ji$_{\rm{CIT}}$, \hi$_{\rm{CIT}}$, and \ki$_{\rm{CIT}}$) using the
following conversion formula \citep{carpenter01}.
\begin{equation}
 J_{\rm{CIT}} = 0.947 J_{\rm{QUIRC}} + 0.053 K_{\rm{QUIRC}} + 0.036
\end{equation}
\begin{equation}
 (J-H)_{\rm{CIT}} = 0.929 (J-H)_{\rm{QUIRC}} + 0.040
\end{equation}
\begin{equation}
 (H-K)_{\rm{CIT}} = 0.975 (H-K)_{\rm{QUIRC}} - 0.027
\end{equation}
The converted \ji-band magnitude, and (\ji--\hi) and (\hi--\ki) colors are given in
Table~\ref{tb:t3}.

\begin{figure}
 \figurenum{8}
 \epsscale{1.0}
 \plotone{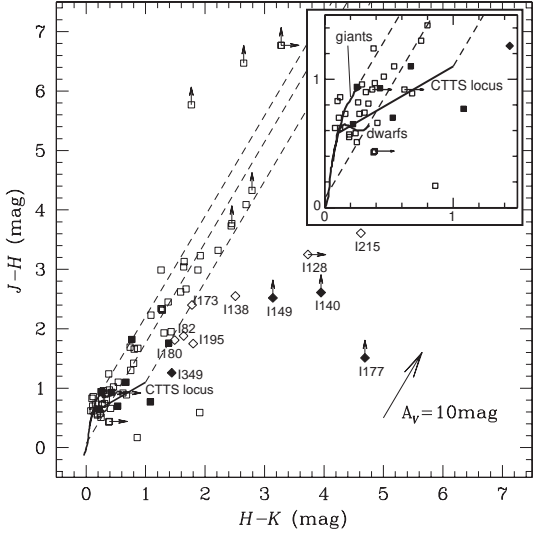}
 \caption{The color-color diagram of QUIRC-IDed \textit{Chandra} sources. Diamonds are
 protostar candidates, which are labeled with the \textit{Chandra} source ID
 (Table~\ref{tb:t3}), while squares are the remainings. Among them, brown dwarf
 candidates are marked filled. The intrinsic colors of dwarfs and giants are given in
 thick solid curves, while the CTTS locus is in thick solid line. The arrow at the
 bottom right gives the reddening vector of $A_{V}=$10~mag. The slope of the reddening
 lines is assumed to be $E(\jh)_{\rm{reddening}}/E(\hk)_{\rm{reddening}}=1.69$
 (Meyer~et~al.~1997). The typical uncertainty is $\sim \pm$0.1~mag. The close-up view
 around the origin is given in the panel at the top right.}\label{fg:f8}
\end{figure}

The color-color diagram was used for the protostar search from QUIRC-IDed
sources. Fig.~\ref{fg:f8} shows the (\ji--\hi) vs. (\hi--\ki) diagram
\citep{lada92}, where the QUIRC-IDed sources are shown in squares. Those with arrows
indicate the lower limit of colors, where the upward arrows are due to the lack of \ji-band
detections and the rightward arrows are due to the saturation in the \ki\ band. The
intrinsic colors of giants and dwarfs are given by the thick solid curves \citep{tokunaga00}.
The emission from the circumstellar disks of the classical T Tauri stars (CTTS) gives a
NIR excess on the giant and dwarf colors, hence aligned from lower-left to upper-right
along the CTTS locus \citep{meyer97}. The color of embedded objects move parallel to the
reddening vector with increasing interstellar or circumstellar gas
(Meyer~et~al.~1997). Therefore, squares between the right and middle dashed lines are
reddened CTTS, while those between the left and middle lines are either reddened
main-sequence stars, weak-line T Tauri stars or some fraction of CTTS. Sources located
to the right of the right dashed line (filled square) may be surrounded by extended
envelopes in addition to the disks, hence are reddened more than the disks alone
\citep{strom95}. Therefore these 11 sources are class~I protostar candidates. We labeled
these sources with $\dagger$ in Table \ref{tb:t3}. 

\begin{figure}
 \figurenum{2}
 \epsscale{1.0}
 \plotone{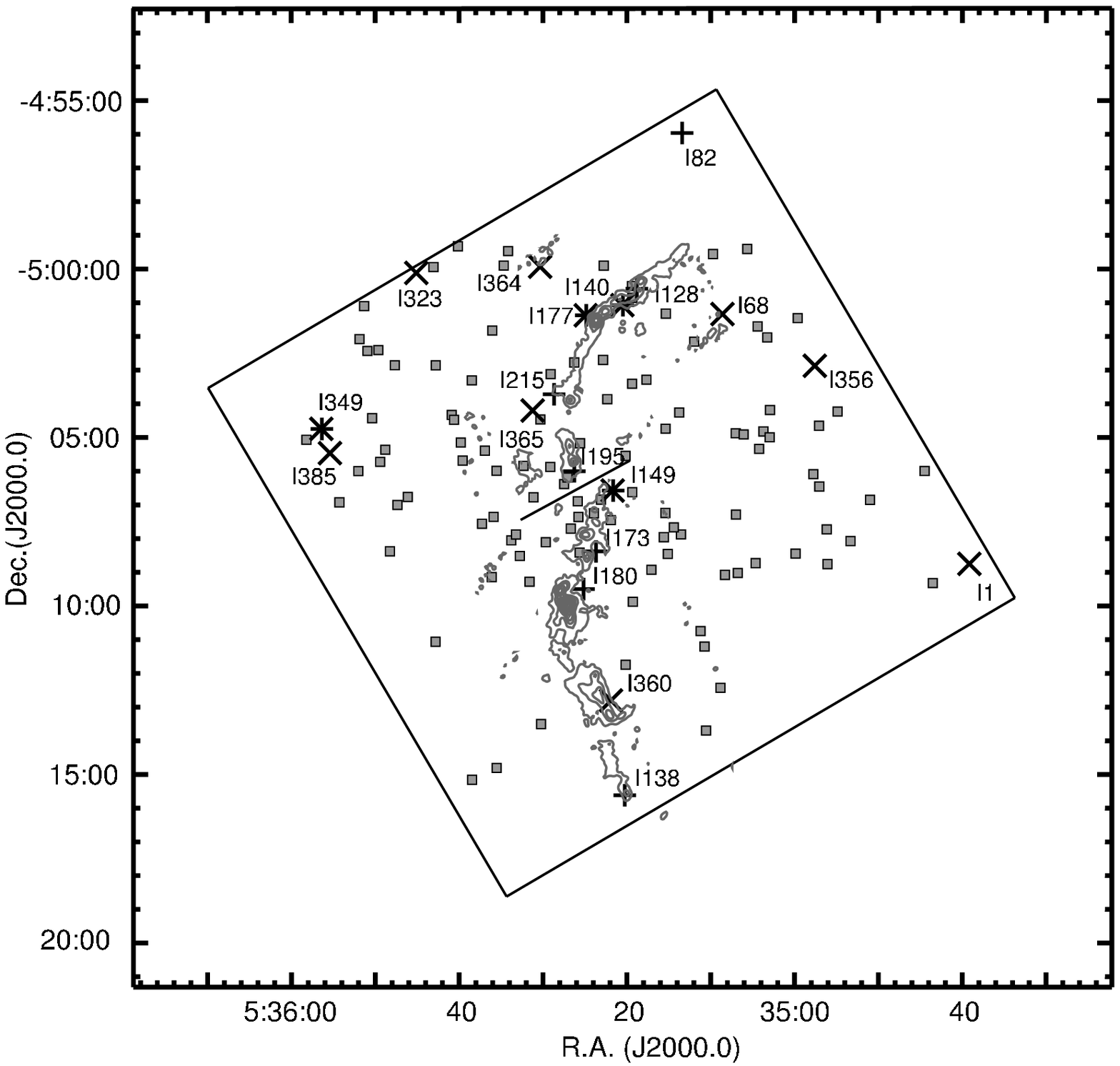}
 \caption{The 1.3~mm intensity map is shown in contours. OMC-2 and OMC-3 are separated
 by the solid line into the southern and northern part of the integral-shaped
 ridge. Squares are positions of \textit{Chandra} sources that have neither 2MASS nor
 QUIRC counterpart, while ``+'' and ``x'' are for the protostar and brown dwarf
 candidates with their \textit{Chandra} IDs (see Sect.\ref{sec:5-2}).}\label{fg:f2}
\end{figure}

Two other features reinforce the idea that they are protostars. The first one is that
many of them (I82, I128, I173, I180, I183, I195, and I215) accompany apparent NIR
nebulosity in the \ki-band image. The second is that their spatial distribution (I128,
I138, I140, I149, I173, I177, I180, I195, and I215) is well correlated with the 1.3~mm
intensity map (Fig.~{\ref{fg:f2}}). Cores seen in the millimeter continuum are the sites
of on-going star formation, and NIR sources along the chain of 1.3~mm cores should be
less evolved YSOs. I82 is located outside of the FOV of 1.3~mm observation
\citep{chini97}, but its protostar nature suggests that the molecular cloud extends
further to the north.

\begin{figure}
 \figurenum{9}
 \epsscale{1.0}
 \plotone{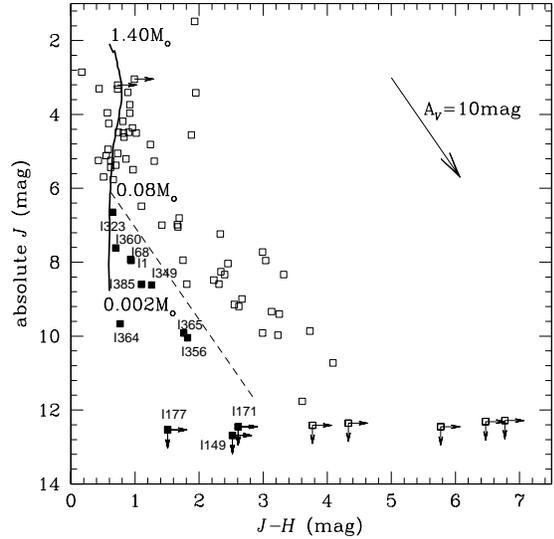}
 \caption{The color-magnitude diagram of QUIRC-IDed \textit{Chandra} sources. Filled are
 brown dwarf candidates, which are labeled with the \textit{Chandra} source ID
 (Table~\ref{tb:t3}). The iso-clone curve in the range of
 1.40M$_{\odot}>$M$>$0.002M$_{\odot}$ at 1~Myr is shown in solid, while the reddening
 line of 0.08M$_{\odot}$ is in dashed line. The arrow at the top left indicates the
 reddening vector of $A_{V}=$10~mag. The typical uncertainty is $\sim \pm$0.1~mag.}\label{fg:f9}
\end{figure}

For the brown dwarf search, we used the \ji\ vs. (\ji--\hi) color-magnitude diagram shown in
Fig.~\ref{fg:f9}. In the same way, the QUIRC-IDed sources are shown as
squares. Squares with the rightward arrows indicate saturation in the \hi\ band, and those
with both the rightward and the downward arrows have \hi- but not \ji-band
detection. The intrinsic colors of stars in the mass range of
0.002M$_{\odot}$--1.40M$_{\odot}$ and age of 1~Myr are shown in the solid curve
\citep{baraffe98}. Due to the extinction, the colors are reddened in the direction of
the reddening vector. Twelve sources (filled squares) are below the reddening line of
0.08M$_{\odot}$, hence these are brown dwarf candidates  with mass less than
0.08M$_{\odot}$. We labeled these sources with $\ddagger$ in Table \ref{tb:t3}. 

As was discussed in the previous section, we have to take back- and fore-ground
contamination into account to discuss the nature of these faint NIR sources. More than
half of these brown dwarf candidate have the \ki-band magnitude of less than 15~mag, where
contamination is not still serious. Some of the sources (I140, I149, I177, and I349)
show NIR excess and some (I68, I140, I149, I177, I360, I364, and I365) are spatially
associated with 1.3~mm cores (Fig.~\ref{fg:f2}), indicating that they are more
likely cloud members and young brown dwarfs \citep{oasa99}.

\subsection{\textit{Chandra} Sources without QUIRC counterpart}

We still have 108 \textit{Chandra} sources that were identified neither by 2MASS nor
QUIRC. This number of unidentified \textit{Chandra} sources is similar with the
\textit{Chandra} studies of other molecular clouds in the Orion--Monoceros complex;
e.g., 101 in Orion nebula cluster \citep{feigelson02} and 87 in Monoceros R2 cloud
\citep{nakajima02}. These X-ray studies have as long exposure time ($\sim$90~ks) as
OMC-2/3 \citep{tsuboi01, tsujimoto02a} and are correlated with the NIR survey data of
the similar depth as this work. The number of unidentified \textit{Chandra} sources can be
consistent with the expected number of background X-ray sources (mostly AGNs)
\citep{mushotzky00,giacconi01}, although the ambiguity due to the extinction is quite
large. Their spatial distribution (squares in Fig.~\ref{fg:f2}) is less correlated with
the 1.3~mm intensity map than the protostar candidates, so they can be mostly background
sources. This does not exclude the possibility of the cloud membership of some
unidentified \textit{Chandra} sources, particularly those associated with the 1.3~mm cores.

\section{Summary}
We conducted mosaic mapping observations of OMC-2/3 in the \ji, \hi, and \ki\ band using
QUIRC.
\begin{enumerate}
 \item We detected 1448 \ki-band sources in $\sim$512 arcmin$^{2}$, for which we
       presented the position, the \ji-, \hi-, and \ki-band magnitude, and the 2MASS
       counterpart. This is the most comprehensive and the deepest NIR source catalog in
       this region.
 \item The astrometric and photometric accuracy of our source list is evaluated using
       the QUIRC--2MASS counterpart pairs to be $\sim$0\farcs2 and
       $\sim$0.2~mag, respectively. The depth of the survey is \ki$\sim$16.0~mag,
       \hi$\sim$16.5~mag, and  \ji$\sim$17.5~mag.
 \item Using the source list, we searched for QUIRC counterpart of 182 2MASS-unIDed
       \textit{Chandra} sources and found 74 of them have QUIRC counterpart.
 \item We examined the nature of the QUIRC-IDed \textit{Chandra} sources with their
       NIR colors, and found 11 protostar and 12 brown dwarf candidates.
 \item The remaining 106 \textit{Chandra} sources with no 2MASS
       nor QUIRC counterpart might be explained by background sources, although some of
       them can be cloud members.
\end{enumerate}

\acknowledgments
The authors express their sincere appreciation to Michael Connelley for taking the
QUIRC calibration data for use, Rolf Chini and Markus Nielbock for providing us the
1.3mm map including unpublished data, and Martin Cohen for letting us use his SKY model
to estimate the galactic star count of our studied field. We also acknowledge the staff
at the University of Hawaii for their support during our observation. M.T. and M.G. are
financially supported by Japan Society for Promotion of Science. This publication makes
use of data products from the Two Micron All Sky Survey, which is a joint project of the
University of Massachusetts and the Infrared Processing and Analysis Center/California
Institute of Technology, funded by the National Aeronautics and Space Administration and
the National Science Foundation.

\clearpage

\clearpage

\clearpage
\begin{deluxetable}{lcl}
 \tabletypesize{\scriptsize}
 \tablecaption{Observation Log\label{tb:t1}}
 \tablecolumns{3}
 \tablewidth{0pt}
 \tablehead{
 \colhead{Date} &
 \colhead{Band} &
 \colhead{Frame} \\
 \colhead{} &
 \colhead{} &
 \colhead{Numbers\tablenotemark{a}}
 }
 \startdata
 2001 Feb 4  & \ki & No.1--No.169\\
 2001 Feb 5  & \ji & No.1--No.52\\
 2001 Feb 6  & \hi & No.1--No.169\\
 2001 Mar 11 & \ji & No.1, No.51--No.169\\
 2001 Mar 12 & \ji & No.1--No.129\\
 2001 Mar 13 & \ji & No.1, No.107--No.169\\
 \enddata
 \tablenotetext{a}{The frame numbers (see Fig.\ref{fg:f1}) covered each night. In the
 \ji-band sweep, for which we spent four nights, some frames are duplicated; i.e., the
 frame No.1 at the beginning of every run to check the telescope positioning accuracy,
 and frame No.51--No.52 and No.107--No.129 to check any difference of image quality in
 different nights. All these duplicated frames were confirmed to have similar quality
 and were combined equally into the final \ji-band mosaic image with the correction for
 the exposure time.}
\end{deluxetable}

\begin{deluxetable}{rccrrrc}
 \tabletypesize{\scriptsize}
 \tablecaption{QUIRC Source List\label{tb:t2}}
 \tablecolumns{8}
 \tablewidth{0pt}
 \tablehead{
 \colhead{ID} &
 \colhead{R.A.\tablenotemark{a}} &
 \colhead{Dec.\tablenotemark{a}} &
 \colhead{\ji\tablenotemark{bcde}} &
 \colhead{\hi\tablenotemark{bcde}} &
 \colhead{\ki\tablenotemark{bde}} &
 \colhead{2MASS} \\
 \colhead{} &
 \colhead{(hh:mm:ss.ss)} &
 \colhead{(dd:mm:ss.s)} &
 \colhead{(mag)} &
 \colhead{(mag)} &
 \colhead{(mag)} &
 \colhead{Counterpart} 
 }
 \startdata
   1 & 05:34:28.94 & $-$05:08:38.7 &    15.52 &    14.13 & 13.74 & 2MASSI J0534289$-$050838\\
   2 & 05:34:29.36 & $-$05:11:14.4 &    17.45 &    16.75 & 16.51 & \\
   3 & 05:34:29.48 & $-$05:05:24.7 &    17.64 &    16.82 & 16.44 & \\
   4 & 05:34:29.60 & $-$05:04:29.0 &    14.05 &    13.47 & 13.15 & 2MASSI J0534295$-$050428\\
   5 & 05:34:29.86 & $-$05:04:05.6 &    12.83 &    12.16 & 11.95 & 2MASSI J0534298$-$050405\\
   6 & 05:34:30.15 & $-$05:11:56.7 &    15.70 &    14.35 & 13.99 & 2MASSI J0534301$-$051156\\
   7 & 05:34:30.27 & $-$05:11:48.2 & 11.53             & 10.64$^{\dagger}$ &  9.95$^{\dagger}$ & 2MASSI J0534302$-$051148\\
   8 & 05:34:30.61 & $-$05:11:05.0 &    13.90 &    13.30 & 13.08 & 2MASSI J0534305$-$051104\\
\enddata
 \tablenotetext{a}{Equinox in J2000.0. The hundredth and the tenth
 order in second of R.A. and Dec. are truncated.}
 \tablenotetext{b}{NIR colors are in the 2MASS color system.}
 \tablenotetext{c}{The upper limit of the flux (the lower limit of the magnitude) is
 given for sources with less than 3$\sigma$ detection in \ji\ and/or \hi\ band and
 is labeled with ``$>$''.}
 \tablenotetext{d}{Saturated QUIRC magnitudes (brighter than 11~mag)
 are replaced with 2MASS magnitudes and labeled $^{\dagger}$ if they
 have the 2MASS counterpart.}
 \tablenotetext{e}{Saturated QUIRC magnitudes (brighter than 11~mag)
 are labeled ``$<$'' if they have no 2MASS counterpart, recognizing
 the value as the lower limit of the flux (the upper limit of the
 magnitude).}
\end{deluxetable}

\begin{deluxetable}{llrrrrr}
 \tabletypesize{\scriptsize}
 \tablecaption{QUIRC-IDed \textit{Chandra} Sources\label{tb:t3}}
 \tablecolumns{7}
 \tablewidth{0pt}
 \tablehead{
 \colhead{X-ray\tablenotemark{abc}} &
 \colhead{QUIRC} &
 \colhead{separation\tablenotemark{d}} &
 \colhead{off-axis\tablenotemark{e}} &
 \colhead{\ji\tablenotemark{f}} &
 \colhead{\ji--\hi\tablenotemark{f}} &
 \colhead{\hi--\ki\tablenotemark{f}} \\
 \colhead{Source ID} &
 \colhead{Source ID} &
 \colhead{(\arcsec)} &
 \colhead{(\arcmin)} &
 \colhead{(mag)} &
 \colhead{(mag)} &
 \colhead{(mag)}
 }
 \startdata
I1$^{\ddagger}$ & TKK~J05343791$-$0508480 & 11.5 & 10.9 & 16.22 & 0.94 & 0.25 \\
I2 & TKK~J05344064$-$0506586 & 3.2 & 10.0 & 15.50 & 2.33 & 1.28 \\
I9 & TKK~J05345056$-$0506383 & 3.4 & 7.6 & 13.96 & 0.51 & 0.25 \\
I46 & TKK~J05350365$-$0513104 & 5.7 & 8.5 & 15.07 & 1.69 & 0.74 \\
I52 & TKK~J05350440$-$0507356 & 1.2 & 4.5 & 14.75 & 1.10 & 0.54 \\
I68$^{\ddagger}$ & TKK~J05350779$-$0501193 & 3.8 & 5.5 & 16.19 & 0.93 & 0.43 \\
I74 & TKK~J05350988$-$0458495 & 0.7 & 7.5 & 16.22 & 3.04 & 1.64 \\
I77 & TKK~J05351053$-$0458460 & 0.6 & 7.5 & $>$20.55 & $>$6.77 & $>$3.28 \\
\enddata
 \tablenotetext{a}{The nomenclature of X-ray sources follow \citet{tsujimoto02a}.}
 \tablenotetext{b}{Protostar candidates are marked with $\dagger$ (see Sect. 5.2.1).}
 \tablenotetext{c}{Brown dwarf candidates are marked with $\ddagger$ (see Sect. 5.2.1).}
 \tablenotetext{d}{The separation between the \textit{Chandra} sources and their QUIRC counterpart.}
 \tablenotetext{e}{The off-axis angle of the \textit{Chandra} sources.}
 \tablenotetext{f}{NIR colors in the CIT color system.}
\end{deluxetable}

\end{document}